\documentstyle[aps,prb,epsf,multicol]{revtex}  
\begin{document}
\draft  
\def\vb{{\bf b}}
\def\vc{{\bf c}}
\def\vd{{\bf d}}
\def\vH{{\bf H}}
\def\vR{{\bf R}}
\def\vm{{\bf m}}
\def\vq{{\bf q}}
\newcommand{\lrule}{ \noindent
  \rule{0.5\textwidth}{0.1mm}\rule{0.1mm}{3pt}\newline }
\newcommand{\rrule}{ \noindent \parbox{\textwidth}{
  \hfill\rule[-3pt]{0.1mm}{3pt}\rule{0.5\textwidth}{0.1mm}}}
%~~~~~~~~~~~~~~~~~~~~~~~~~~~~~~~~~~~~~~~~~~~~~~~~~~~~~~~~~~~~~~~~~~~~~~~~~~~~~~~
\title{Magnetic Field Effects on Neutron Diffraction in the Antiferromagnetic 
	Phase of $UPt_3$}
\author{Juana Moreno and J. A. Sauls}  
\address{Department of Physics \& Astronomy \\
	 Northwestern University, Evanston IL 60208}  
\date{\today}  
\maketitle
%~~~~~~~~~~~~~~~~~~~~~~~~~~~~~~~~~~~~~~~~~~~~~~~~~~~~~~~~~~~~~~~~~~~~~~~~~~~~~~~
\begin{abstract} 
We discuss possible magnetic structures in UPt$_3$ based on our
analysis of elastic neutron-scattering experiments
in high magnetic fields at temperatures $T<T_N$. The existing 
experimental data can be explained by a single-{\bf q} 
antiferromagnetic structure with three independent domains. 
For modest in-plane spin-orbit interactions, the Zeeman 
coupling between the antiferromagnetic order parameter and 
the magnetic field induces a rotation of the magnetic 
moments, but not an adjustment of the propagation vector of 
the magnetic order. A triple-{\bf q} magnetic structure is 
also consistent with neutron experiments, but in general leads to
a non-uniform magnetization in the crystal. New experiments 
could decide between these structures.
\end{abstract}  
\begin{multicols}{2}

The coexistence of antiferromagnetic and superconducting order for five
of the six heavy fermion superconductors suggests a deep connection 
between these two aspects of heavy fermion physics. In these materials 
the f-electrons are involved in the superconducting transition, just as 
they are in the formation of the coherent heavy fermion band, but their
precise role in the development of the unconventional superconducting phase 
is still unclear.

The magnetic field versus temperature phase diagram of $UPt_3$ provided
compelling evidence of unconventional superconductivity
in U-based heavy fermion materials.\cite{has89,ade90,bru90} 
In order to explain the phase diagram of UPt$_3$ several authors
proposed a multicomponent order parameter based on a multi-dimensional 
representation of the hexagonal point group.\cite{hes89,mac89,joy92,sau94,mac98}
In these models a weak symmetry breaking field (SBF) is invoked.
This SBF lifts 
the  degeneracy of the multi-dimensional representation and leads to multiple 
transitions at lower temperatures and higher fields (see also the reviews
in Ref. \onlinecite{hef96}).

A natural candidate for the role of SBF is the weak antiferromagnetic
order shown by neutron scattering measurements below 
$T_N=6 K$.\cite{aep88,isa95,kei99a} The ordered moment is unusually small, 
only $0.02 \mu_B$ per U atom, and is directed
in the basal plane, thus breaking the in-plane hexagonal symmetry. 
Evidence in support of an antiferromagnetic SBF coupled to the
the superconducting order parameter is based on the correlation
between changes in the magnitude of the ordered moment 
and the splitting of the double transition. Both the splitting and the
AFM order parameter are suppressed under applied pressure of 
$p_c\approx 3.5\,\mbox{kbar}$.\cite{tra91,hay92}
The effect of $Pd$ is the opposite; the splitting and the ordered 
moment increase with increasing $Pd$ substitution.\cite{kei99b}

Most thermodynamic and transport measurements have failed to detect
a signature of AFM ordering near
$T_N\simeq 6\,\mbox{K}$.\cite{fis91,tou96,kei99c,dal95} 
However,
evidence of magnetic ordering is observed to onset at $T_N$ in the
magnetoresistance.\cite{beh90}
The transition has other unusual characteristics as well,
including finite range correlations, $\xi_{AFM}\sim 300-500$ \AA,
depending on the crystalline direction and sample. By contrast ,
$(U,Th)(Pd,Pt)_3$ alloys exhibit AFM ordering at $T_N\approx
6\,\mbox{K}$, but with ordered moments of conventional size, $\mu\sim
0.65\mu_B/U$-ion, and resolution-limited Bragg peaks at the same positions 
as pure $UPt_3$.\cite{gol86,fri87} 
Based on these facts, several authors have argued that the 
anomaly at $6 K$ does not indicate the onset of true long range magnetic 
ordering but finite-range AFM correlations,\cite{fom96} which may also be
fluctuating on time scales of order 
$5 \cdot 10^{-10}$ s to $10^{-7}$ s.\cite{oku98} 

Given the uncertainties about the nature of magnetic order
in UPt$_3$, studies of the field dependence of the magnetic order
were performed in order to help clarify these issues. Two experimental 
groups have measured neutron scattering ratios in magnetic fields up to
$3.5\,\mbox{T}$\cite{lus96a} and $12 T$.\cite{dij98}
Both studies concluded that applied magnetic fields have no effect
on the magnetic order of $UPt_3$, whether it be in aligning  the moments 
or in domain selection. Our analysis and interpretation of
these experiments leads to the conclusion that there is still room 
for a conventional dependency on the magnetic field and that additional 
neutron scattering data is necessary to clarify this issue.

\begin{figure}[h]
\begin{minipage}{\linewidth}
\centerline{\epsfxsize8.5cm\epsffile{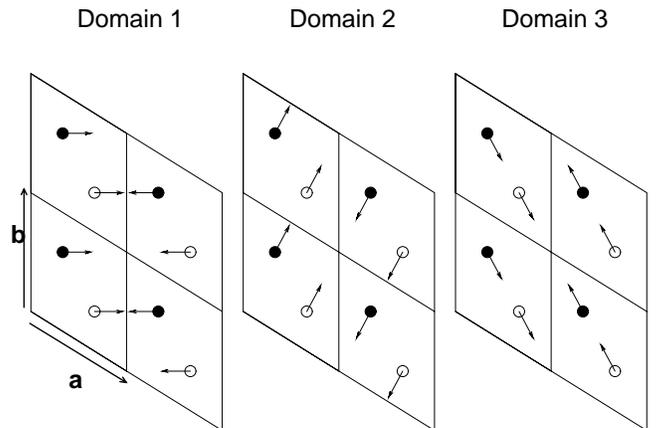}}
\caption[]
{The three equivalent domains for the configuration with propagation vector 
$\vec{q_1}=\vec{a}^*_1/2$. The other two configurations 
($\vec{q_2}=\vec{a}^*_2/2,\vec{q_3}= (\vec{a}^*_1-\vec{a}^*_2)/2$) also 
present identical domain structures. Black filled circles represent
U atoms in the $z=c/4$ plane, empty circles represent U atoms
on the  $z=3c/4$ plane.}
\label{fig:domains} 
\end{minipage}
\end{figure}

We start from the conventional
assumption of tiny antiferromagnetically ordered moments at each U site.
These moments ($\vec{m}$) are assumed to lie on the basal plane 
due to a strong uniaxial anisotropy arising from spin-orbit coupling. 
In addition, there is an in-plane (hexagonal) anisotropy energy
which favors alignment of the moments along any of the
three directions perpendicular to the hexagonal lattice vectors 
(Fig. \ref{fig:domains}).

Neutron-scattering and x-ray experiments\cite{aep88,isa95,kei99a} 
show antiferromagnetic order with three possible propagation vectors: 
$\vec{q}_1=\vec{a}^*_1/2, \vec{q}_2=\vec{a}^*_2/2,
\vec{q}_3=(\vec{a}^*_1-\vec{a}^*_2)/2$, where 
$\vec{a}^*_1=\frac{4\pi}{\sqrt{3} a}(1,0,0)$, 
$ \vec{a}^*_2= \frac{4\pi}{\sqrt{3} a}(1/2,\sqrt{3}/2,0)$ 
and $ \vec{a}^*_3=\frac{2\pi}{c}(0,0,1)$ are the reciprocal 
vectors of the hexagonal lattice with dimensions 
$a=5.74$ \AA \hspace{0.05in} and $c=4.89$ \AA. 
The two U moments in each crystallographic unit 
cell have to align ferromagnetically in order to account for 
most of the zero-intensity Bragg points in the diffraction pattern.
But, in general, the magnetic structure cannot be fully determined 
by standard neutron diffraction experiments, since these experiments
provide information only about the Fourier components of the magnetic
moment. Single- and multi-{\bf q} magnetic structures display
the same magnetic Bragg peaks, and cannot be distinguished 
unless uniaxial stress or a magnetic field is applied.\cite{ros84}

The magnetic neutron scattering rate per solid angle 
is proportional to\cite{ros84,squ78}
\begin{equation}
\bigg(\frac {d\sigma}{d\Omega}\bigg )_{\vec{Q}} 
\propto \sum_{\vec{Q}_m}  |F_{M \perp}(\vec{Q})|^2 
\delta(\vec{Q}-\vec{Q}_m) 
\end{equation}
where $\vec Q$ is the momentum transfer, $\vec{Q}_m$ are the momenta 
of the magnetic Brag peaks and $F_{M \perp}(\vec{Q})$ is the component of the 
magnetic  structure factor perpendicular to the momentum transfer.
We can define the magnetic  structure factor as
\begin{equation}
\vec{F}_M(\vec{Q})= \frac{1}{N} \sum_{n,j} \vec{m}_{nj}
f_{nj}(\vec{Q}) e^{i\vec{Q}\cdot\vec{R}_{nj} - W_j}
\end{equation}
where $\vec{m}_{nj}$ is the magnetic moment of the $j^{\mbox{th}}$ ion in the 
$n^{\mbox{th}}$ unit cell, $f_{nj}$ is its atomic form factor, 
$\vec{R}_{nj}$ is its position and $W_j$ is the Debye-Waller factor. 

The spatial distribution of magnetic moments can be Fourier expanded as
$\vec{m}_{n,j}=\sum_{\vec{q}} \vec{m}_{\vec{q},j} e^{-i \vec{q}\cdot\vec{R}_n}$,
where the form factor associated with this multi-{\bf q} magnetic structure is
$\vec{F}_{M}(\vec{Q}=\vec{Q}_{nm}+\vec{q})=
\sum_{j} \vec{m}_{\vec{q},j} f_{j}(\vec{Q}) e^{i\vec{Q}\cdot\vec{r}_j -W_j}$
where $\vec{r}_j$ are the positions of the magnetic ions in the
unit cell and $\vec{Q}_{nm}$ label the reciprocal lattice vectors.
Thus, in a  material with only one type of magnetic ion
the scattering rate becomes
\end{multicols}
\lrule
\begin{equation}
\bigg(\frac {d\sigma}{d\Omega}\bigg )_{\vec{Q}} 
\propto \sum_{\vec{Q}_{nm},\vec{q}}  [1-(\hat{Q} \cdot \hat{m}_{\vec{q}})^2]
|f(\vec{Q})|^2 \bigg{|}\sum_{\vec{r}_i}e^{i \vec{Q} \cdot \vec{r}_i}
\vec{m}_{\vec{q}} \bigg | ^2
\delta(\vec{Q}-(\vec{Q}_{nm}+\vec{q}))
\,.
\end{equation}
\rrule
\begin{multicols}{2}

\noindent Thus, the $UPt_3$ diffraction pattern can either be associated with a
triple-{\bf q} structure where $\vec{q}_1$, $\vec{q}_2$ and $\vec{q}_3$ are
present at each uranium site or with a single-{\bf q} structure where separate
regions of the crystal will order with different propagation vectors.
It has been inferred from the fact that there is no intensity at the
$\vec{q}_1=[1/2,0,0]$ position that the magnetic moment lies parallel to its
propagation vector.\cite{gol86,fri87} 
This is  the case in the U-monochalcogenides and
U-monopnictides with cubic NaCl structure, which order with
magnetic moments $\mu \simeq 1-3\, \mu_B$.\cite{ros84}
A moment directed along $\vec{q}$ would
also occur for a triple-{\bf q} structure, but
it is not clear that this condition must be fulfilled in
the single-{\bf q} structure. The intensity of 
$\vec{q}_2=[0,1/2,0]$ and $\vec{q}_3=[1/2,-1/2,0]$ peaks 
has not been reported for UPt$_3$. It is possible that
the sample preparation methods make domain ``1'' (Fig. \ref{fig:domains})
preferable over domains ``2'' and ``3''. However, measuring the
intensity of these three peaks in the same single crystal would
allow one to determine if the magnetic moments do lie parallel to the
propagation vector of the domain.

Below we discuss the field dependence of the magnetic neutron scattering
intensity for the possible magnetic structures. We first discuss
the field dependence of single-{\bf q} structures, then we
comment  on the possibility of a triple-{\bf q} magnetic  structure.
The magnetic unit cell of a single-{\bf q} structure results from 
doubling the hexagonal unit cell along one in-plane direction, 
reducing the hexagonal symmetry to orthorhombic.

Transmission electron microscope images provide direct
observation of basal plane, as well as prism plane,
stacking faults in pure single crystals.\cite{hon00}
These defects are observed even in the crystals with
the highest residual resistance ratios. We hypothesize 
that these defects pin AFM domain walls in the ab-plane and
fix the spatial distribution of domains.\cite{ama98,hux00}
%~~~~~~~~~~~~~~~~~~~~~~~~~~~~~~~~~~~~~~~~~~~~~~~~~~~~~~~~~~~~~~~~~~~~~~~~~~~~~~~
%~~~~~~~~~~~~~~~~~~~~~~~~~~~~~~~~~~~~~~~~~~~~~~~~~~~~~~~~~~~~~~~~~~~~~~~~~~~~~~~

In an antiferromagnet 
the Zeeman energy prefers the staggered magnetization
to be perpendicular to the field. Thus, a sufficiently strong
magnetic field applied in the hexagonal plane will give rise to domain 
reorientation by overcoming the in-plane anisotropy energy.
The magnitude of the staggered magnetization will remain roughly the same,
modulated only by a small in-plane anisotropy energy.\cite{sau96}
Therefore, for a given magnetic Bragg peak,
the ratio between the scattering rate
at high field and at zero field is\cite{kit94}
\begin{equation}
r=\frac{d\sigma/d\Omega|_{H\rightarrow\infty}}{d\sigma/d\Omega|_{H=0}}
\approx\frac{\langle  1-(\hat{Q}\cdot\hat{m}_{H\rightarrow\infty})^2\rangle}
{\langle  1-(\hat{Q} \cdot \hat{m}_{H=0})^2 \rangle},
\label{eq:ratioH} 
\end{equation}
where $\langle ... \rangle$ refers to an average over domains.

%~~~~~~~~~~~~~~~~~~~~~~~~~~~~~~~~~~~~~~~~~~~~~~~~~~~~~~~~~~~~~~~~~~~~~~~~~~~~~~~
\begin{figure}
\begin{minipage}{\linewidth}
\centerline{\epsfxsize8.5cm\epsffile{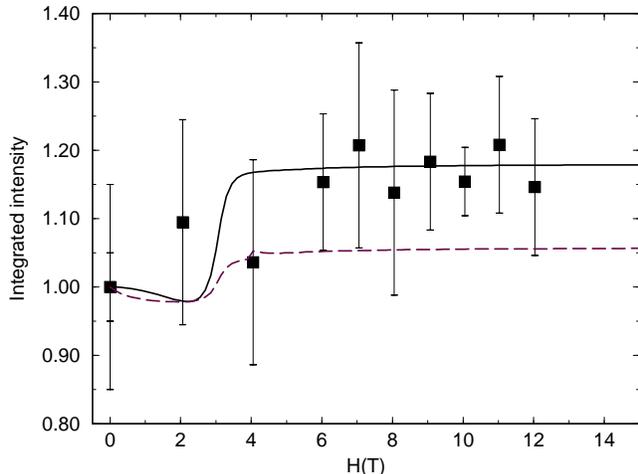}}
\caption[]
{Relative integrated intensity of the magnetic Bragg peak  
$\vec Q=[1/2,0,1]$ as a function of applied magnetic field. 
H is parallel to the ``a'' axis. The solid line corresponds to a crystal
with only domain ``1''populated, and the dashed line represents a 
sample with three equally populated domains. The parameters we  used 
(refer to Eq. (\ref{eq:free})) are: $H_{an}=1.5T$, $U_{an}=0.02 U_{ex}$ 
and $r_{st}=0.02$. The calculated curves are compared with measurements
of van Dijk et al.
\cite{dij98}(black squares).} 
\label{fig:neutronH} 
\end{minipage}
\end{figure}
%~~~~~~~~~~~~~~~~~~~~~~~~~~~~~~~~~~~~~~~~~~~~~~~~~~~~~~~~~~~~~~~~~~~~~~~~~~~~~~~

Let us analyze the experimental data based on Eq. (\ref{eq:ratioH}).
The staggered magnetization lies on the basal plane,
$\hat{m}=(\cos\theta,\sin\theta)$. Van Dijk et al. \cite{dij98} chose
a configuration with H parallel to the ``a'' axis ($\theta_H=-30^o$ in 
Eq. \ref{eq:free}) and a momentum transfer 
$\vec Q=[1/2,0,1]=2\pi ((1/\sqrt{3}a),0,(1/c))$, which gives
$\hat{Q}=(0.441,0,0.897)$ 
and
\begin{equation}
r=\frac 
{ 1-(0.441\,\hspace{0.01in}\cos(\theta_H+\pi/2))^2}
{\langle  1-(0.441\,\hspace{0.01in}\cos(\theta))^2\rangle}= 1.05
\end{equation}
for three equally populated magnetic domains. This ratio can be increased to
$r=1.18$ by assuming that only the domain with the staggered magnetization
parallel to the propagation vector is populated 
(domain ``1'' in Fig. (\ref{fig:domains})).
Thus, even in the case of complete domain reorientation,
the neutron scattering rate at $\vec Q=[1/2,0,1]$ 
in high fields can increase at most by 18\% 
over its value at zero field.
Figure \ref{fig:neutronH} shows the experimental data and the 
theoretical curves for a model with equally populated domains and for 
a model with only  domain ``1'' populated. Although the theoretical
calculation associated with domain ``1'' is in good agreement with 
the data, it is not possible to conclude whether or not the U 
moments rotate with the field because of the small change in intensity 
that is expected for this Bragg peak and the large error
bars that are reported for the intensity. Note that 
the error bars for this measurement are comparable to the maximum  
change in the intensity ratio. In our calculation
we have assumed an anisotropy field of $H_{\mbox{\small an}}=1.5\,T$.
However, much smaller values are consistent with the limited data.
The precise value of the additional parameters in our model play a role 
only in the region of small magnetic fields. For fields 
$H > 2 H_{an}$ the ratio between the intensity at high fields 
and at zero field saturates at its upper limit, which 
is determined by purely geometrical arguments.

Earlier analysis \cite{dij98} was based on the assumption that the staggered 
magnetic moment is {\it always} parallel to its propagation vector.
Thus, it was expected that a sufficiently high magnetic field 
parallel to the ``a'' axis would select domain ``2'' 
with propagation vector  $\vec{q}_2$ throughout the sample. 
As a consequence, the magnetic intensity at 
$\vec Q=[1/2,0,1]=\vec{q}_1+[0,0,1]$ 
was expected to drop to zero.
However, as we show in Fig. \ref{fig:neutronH}, if we assume that
the spatial distribution of domain walls is pinned,
the form factors for $\vec Q =[1/2,0,1]$, which is a vector mostly out of 
the hexagonal plane, lead to a much smaller variation of 
the intensity with the field.

%~~~~~~~~~~~~~~~~~~~~~~~~~~~~~~~~~~~~~~~~~~~~~~~~~~~~~~~~~~~~~~~~~~~~~~~~~~~~~~~
\begin{figure}
\begin{minipage}{\linewidth}
\centerline{\epsfxsize8.5cm\epsffile{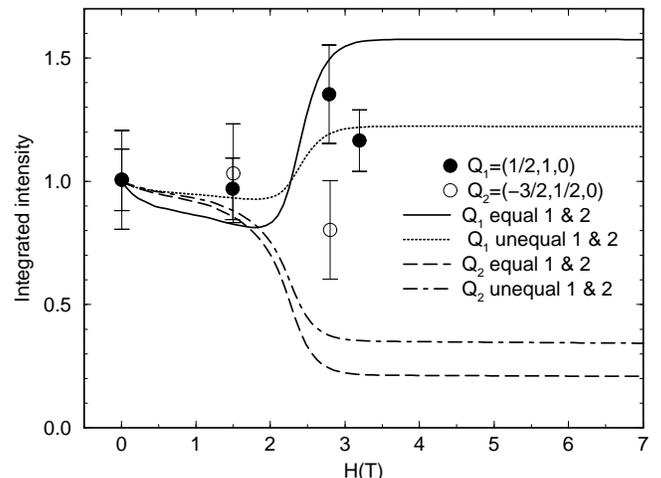}}
\caption[]
{Normalized integrated scattering intensity as a function of the field for 
$\vec{Q}_1=[1/2,1,0]$ and $\vec{Q}_2=[-3/2,1/2,0]$.
The magnetic field points along the b axis.
Calculated curves are compared with measurements 
of  Lussier et al.\cite{lus96a} 
We show calculations for two domain structures: domain 
``1'' and ``2'' equally populated  and domain 
``1'' on $3/4$ of the sample, domain ``2'' on $1/4$ . We
used the same parameters as those for Fig. (\ref{fig:neutronH}):
$H_{an}=1.5T$, $U_{an}=0.02 U_{ex}$ and $r_{st}=0.02$.}  
\label{fig:taillefer} 
\end{minipage}
\end{figure}
%~~~~~~~~~~~~~~~~~~~~~~~~~~~~~~~~~~~~~~~~~~~~~~~~~~~~~~~~~~~~~~~~~~~~~~~~~~~~~~~

Larger expected ratios between the low- and high-field 
intensities are obtained 
with the experimental setup used by Lussier et al.\cite{lus96a}
They measured the neutron scattering cross-section at three different
momentum transfers, all in the basal plane:
$\vec{Q}_1=[1/2,1,0],\vec{Q}_2=[-3/2,1/2,0]$ and $\vec{Q}_3=[-1,3/2,0]$.
The magnetic field was oriented along the $\vb$ axis.
Lussier et al.\cite{lus96a} report data for $\vec{Q}_1$ and $\vec{Q}_2$, 
and magnetic fields up to $3.5T$.
We can estimate from Eq. (\ref{eq:ratioH})
the ratio between high- and zero-field intensity
for any distribution of domains in the crystal.
A crystal with equally populated domains will display
the following ratios for the neutron scattering rate 
at high fields and zero field:
\begin{equation}
r(\vec{Q}_1)=0.86\,,
\hspace{0.2in} r(\vec{Q}_2)=0.21\,,
\hspace{0.2in} r(\vec{Q}_3)=1.93
\,.
\end{equation}
If domain 3 is unpopulated and  domains 1 and 2 are equally populated 
the ratios should be:
\begin{equation}
r(\vec{Q}_1)=1.60\,,
\hspace{0.2in} r(\vec{Q}_2)=0.20\,,
\hspace{0.2in} r(\vec{Q}_3)=1.38\,,
\end{equation}
and if only the domain with the magnetization parallel to
the propagation vector is occupied (e.g. domain ``1'' for $\vec{Q}_1$)
then,
\begin{equation}
r(\vec{Q}_1)=1\,,
\hspace{0.2in} r(\vec{Q}_2)=0.25\,,
\hspace{0.2in} r(\vec{Q}_3)=2.25\,.
\end{equation}

Figure \ref{fig:taillefer} displays the experimental data of Ref.
\onlinecite{lus96a}, and theoretical calculations for two samples,
one with domains ``1'' and ``2'' equally populated at zero field,
another with domains ``1'' and ``2'' unequally populated.
The parameters of the model are the same ones used to fit
the data at $\vec{Q}=[1/2,0,1]$ in Fig. \ref{fig:neutronH}.
We conclude that the limited data for $\vec{Q}_1$ and $\vec{Q}_2$
is roughly consistent with either one or two unequally populated domains,
particularly if $H_{an}\agt 2.5\,\mbox{T}$.
Previous analysis of these results was also based on the assumption
that the propagation vector of the magnetic domains follows the 
rotation of the magnetic moments.\cite{lus96a} 
Thus, at high fields it was expected 
that the intensity of the $\vec{Q}_2$ and  $\vec{Q}_3$ peaks would
be suppressed to zero, while increasing the intensity of the
$\vec{Q}_1$ peak to roughly three times its zero field value.

The theoretical curves have been calculated using 
the free energy functional,\cite{and80,sau96}
\end{multicols}
\lrule
\begin{eqnarray}
\bar{F}_{AFM}=-2 (1-\bar{T})|\vec{m}_0|^2+|\vec{m}_0|^4 +
\bar{U}_{an}|\vec{m}_0|^6 (r_{6}-\cos(6 \theta))
\label{eq:free}
+\bar{U}_{an} \bar{H}^2|\vec{m}_0|^2 \cos^2(\theta-\theta_H))+\\
+r_{D} \bar{U}_{an} \bar{H} |\vec{m_0}| |\sin(\theta-\theta_H))| +
 r_{st} |\vec{m}_0|^2\Bigg(
\bigg(\frac {\partial(\cos(\theta))}{\partial H}\bigg)^2 +
\bigg(\frac {\partial(\sin(\theta))}{\partial H}\bigg)^2 \bigg)\nonumber
\end{eqnarray}
\rrule
\begin{multicols}{2}
\noindent where all energies are measured in units of the exchange energy, 
$U_{\mbox{\small ex}}$, which is defined as the absolute value of the free
energy at zero temperature and field in the absence of any anisotropy
energy. The magnetic order  parameter is restricted to the basal plane by the
large uniaxial anisotropy energy (not shown in Eq. \ref{eq:free})
and it is measured with respect to the antiferromagnetic order parameter
in the exchange approximation:
$\vec{m}_0=\vec{m}/|\vec{m}_{ex}|=|\vec{m}_0|(\cos \theta,\sin\theta,0)$. 
The renormalized temperature is defined as $\bar{T}=T/T_N$, with
$T_N$ as the N{\'e}el temperature. 
The magnetic field  $\bar{H}$ is measured in units of the in-plane
anisotropy field, $H_{an}$. 
The first two terms of the free energy correspond to the exchange 
energy. For $\bar{T}< 1$ antiferromagnetic order with magnetic moment 
$|\vec{m}_0|=|\vec{m}|/|\vec{m}_{ex}|=\sqrt{1-\bar{T}}$ and 
free energy $\bar{F}_{AFM}=F_{AFM}/U_{ex}= - (1-\bar{T})^2$ is stable.
The sixth-order term is the leading term in the in-plane anisotropy energy;
it favors alignment along the three directions perpendicular to the 
hexagonal lattice vectors: $\theta=n(\pi/3)$, where $n$ is an integer.
The in-plane anisotropy energy induces a hexagonal modulation of the
upper critical field as a function of the orientation of the field in 
the basal plane.\cite{kel95} From the magnitude of this hexagonal modulation
we estimate an anisotropy energy of 
$\bar{U}_{an}=U_{an}/U_{ex}\sim 0.02$.\cite{sau96}
The parameter $r_{6}$ must be bigger than one in order to have a
stable free energy. We use $r_{6}=1.5$ in our calculations, however, its
precise value does not play any significant role in the minimization
of the free energy.

The fourth term is the Zeeman energy for an antiferromagnet,
$F_{Z}=g (\vec{m} \cdot \vec{H})^2$, which is quadratic in $H$ and favors 
perpendicular alignment ($g>0$) of the staggered moment and the magnetic field. This
term can be written in the form,
\begin{equation}
F_{Z}=\frac{{U}_{an}}{U_{ex}} 
\Big(\frac{H}{H_{an}}\Big)^2 \Big(\frac{\vec{m}}{|\vec{m}_{ex}|}\Big)^2 
\cos^2(\theta-\theta_H)
\,,
\end{equation}
where $H_{an}=(1/|\vec{m}_{ex}|)\sqrt{{U}_{an}/(g U_{ex})}$
and $\theta_H$ is the angle of the magnetic field with the 
$\vec{a}^*_1$ reciprocal vector.

The fifth term in Eq. \ref{eq:free} is the Dzyaloshinskii-Moriya term 
describing the {\sl linear} coupling of the sublattice magnetization to
the magnetic field, $F_D=g'\vd\cdot\left(\vH\times\vm_0\right)$. 
This term corresponds to the Zeeman coupling of a weak ferromagnetic
(FM) moment in systems which are predominantly antiferromagnetic.
Its origin is the anisotropic superexchange coupling between magnetic 
moments, $\sim\vec{D}_{ij} \cdot \vec{S}_i\times \vec{S}_j$, where
$\vec{D}_{ij}$ are the Moriya vectors for different bonds on the lattice,
and which are related to each other by lattice symmetries.\cite{mor60,she92}
In the case of $UPt_3$, $\vec{D}_{ij}=0$ when $i$ and $j$ are 
nearest-neighbor U sites, while $\vec{D}_{ij}=\pm  |d| \hat{c}$,
independent of the direction of the staggered magnetic moment, when 
$i$ and $j$ refer to next-nearest-neighbor U atoms.\cite{superexchange}
This superexchange coupling generates the Dzyaloshinskii
term in the free energy which can be expressed as 
$\bar{F}_D= r_D\bar{U}_{an}\bar{H}|\vec{m_0}| |\sin(\theta-\theta_H))|$
shown in Eq. \ref{eq:free}.

\begin{figure}
\begin{minipage}{\linewidth}
\centerline{\epsfxsize8.5cm
\epsffile{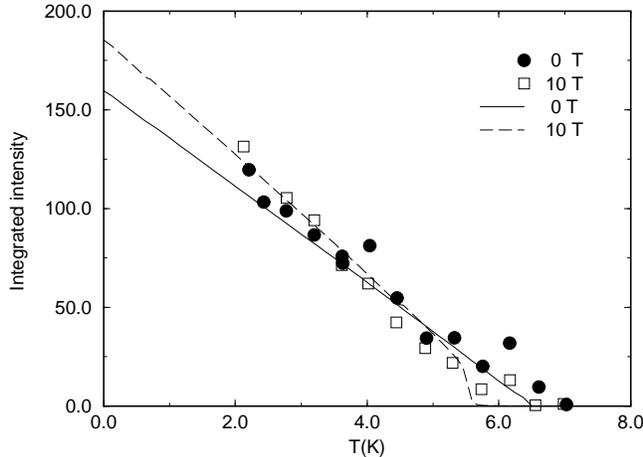}}
\caption[]
{Temperature dependence of the integrated intensity of the magnetic
Bragg peak  $\vec Q=[1/2,0,1]$ in a magnetic field of $H=0$ and $10\,T$.
The solid line represents a calculation at $H=0\,T$, and the dashed line 
shows the dependence at $H=10\,T$. The calculations assume that only
domain ``1'' is populated, and we have used the same parameters
as those used for the calculations shown in Fig. (\ref{fig:neutronH}) 
plus a weak ferromagnetic coupling proportional to $r_D=0.5$.
The experimental data is that reported by van Dijk et al. 
\cite{dij98} at zero field (black circles) and at $10\,T$ (white squares).}
\label{fig:neutronT} 
\end{minipage}
\end{figure}

For low temperatures the effect of the Dzyaloshinskii-Moriya term is to 
generate a tiny ferromagnetic moment at the price of a small reduction
in the magnitude of the staggered moment. However, for temperatures close
to $T_N$, the Dzyaloshinskii-Moriya energy is comparable to the exchange
energy, and leads to a significant reduction in the magnitude of the AFM 
moment and, as a consequence, the intensity of the magnetic Bragg peaks. 
We can define a crossover temperature in terms of the parameters of the free
energy, $\bar{T}_D= 1-\sqrt[3]{r_{D}^2 U^2_{an}\bar{H}^2}$. Although the
staggered moment vanishes precisely at the N{\'e}el temperature,
for $\bar{T}_D < \bar{T} < 1$ the moment decreases rapidly before the 
transition at $\bar{T}=1$. Thus, $\bar{T}_D$ could be mis-identified 
as the N{\'e}el temperature of the sample. The 
Dzyaloshinskii-Moriya term provides an explanation for
the crossing of the intensity curves for zero and high fields as a
function of temperature as shown in Fig. \ref{fig:neutronT}.

The Dzyaloshinskii-Moriya coupling also provides an explanation for
the linear term in the field dependence of the
magnetoresistance,\cite{beh90} which onsets at the Ne\`el transition
and increases for $T < T_N$. It has been shown that 
a linear term in the transverse magnetoresistance 
is present in antiferromagnetic structures admitting the
existence of weak ferromagnetism.\cite{tur63} 
Indeed it follows from Onsager relations for the resistivity
that a magnetoresistance which is linear in field in a AFM 
requires the Dzyaloshinskii-Moriya coupling.

%~~~~~~~~~~~~~~~~~~~~~~~~~~~~~~~~~~~~~~~~~~~~~~~~~~~~~~~~~~~~~~~~~~~~~~~~~~~~~~~
%~~~~~~~~~~~~~~~~~~~~~~~~~~~~~~~~~~~~~~~~~~~~~~~~~~~~~~~~~~~~~~~~~~~~~~~~~~~~~~~

Finally, the last term in Eq. \ref{eq:free} describes the ``stiffness'' 
of the order parameter with respect to rotations in the ab-plane.
This stiffness originates from the formation of domains
in which the staggered moment points in the same direction within
each domain. An inhomogeneous domain structure gives rise to domain
walls separating differently oriented domains. The energy associated with
the domain wall is obtained from the gradient energy,
$\kappa_{ijkl} (\partial m_j/\partial x_i)(\partial m_l/\partial x_k)$,
which must be included in the free energy functional.
For an individual domain wall, the gradient energy can be written 
as an integral over the domain wall surface $\Omega$,\cite{ast90}
\begin{equation}
F_{\mbox{wall}}\propto
\int_{\Omega} d\Omega \int^{\sigma_2}_{\sigma_1}\Big[ \bigg(  
\frac{\partial {\bf\hat{m}_x}}{\partial\sigma}\bigg)^2 + 
\bigg(\frac{\partial {\bf\hat{m}_y}}{\partial\sigma}\bigg)^2\Big] d\sigma
\end{equation}
where $\sigma$ is the coordinate perpendicular at each point
to the wall surface. The width of the wall is given by 
$\sigma_2-\sigma_1$ and ${\bf\hat{m}_x}$, ${\bf\hat{m}_y}$ are 
the components  of the unit vector ${\bf\hat{m}}=\vm/|\vm|$. This 
unit vector satisfies the boundary conditions,
${\bf\hat{m}}(\sigma_2)={\bf\hat{m}}_{eq}(H+\Delta H)$
and ${\bf\hat{m}}(\sigma_1) ={\bf\hat{m}}_{eq}(H)$, 
where ${\bf\hat{m}}_{eq}(H)$ is the equilibrium orientation of the
staggered magnetic moment in the presence of a magnetic field $\vec{H}$.
In quasiequilibrium the direction of the magnetic moment evolves 
smoothly through the domain wall between its values corresponding to 
different equilibrium field orientations, 
${\bf\hat{m}}_{eq}(H+\Delta H)$ and ${\bf \hat{m}}_{eq}(H)$.
By scaling the width of the domain wall to $\Delta H$ 
we obtain the stiffness energy in the form of the last
term in Eq. \ref{eq:free}.

The stiffness energy is important in the region of intermediate fields,
where the normalized neutron intensity increases from a
value close to the one at zero field to its value at high fields.
The initial drop of the neutron intensity as a function of the applied field
(Fig. (\ref{fig:neutronH}) and (\ref{fig:taillefer})) is a combined effect
of the anisotropy and stiffness energies. This drop is due 
to an initial reduction of the magnitude of the magnetic moment.
Small fields do not induce rotation; instead the magnitude of the staggered
moment is reduced. Higher fields are able to rotate the moments by 
overcoming the anisotropy and stiffness energies. Consequently, the
Zeeman energy is reduced to zero and the rotated moment recovers
its value at zero field.

So far we have discussed single-$\vq$ structures or multi-domain
single-$\vq$ structures. Triple-{\bf q} structures are also possible. 
By symmetry each component, $\vec{m}_{\vec{q}_i}$, has the same amplitude. 
Triple-{\bf q} antiferromagnetic order occurs in the
$NaCl$-type monopnictide $USb$,\cite{ros84} 
in the $CsCl$-type $DyAg$ \cite{mor89} and $NdZn$,\cite{mor75,mor74} 
and in the $AuCu_3$-type $TmGa_3$.\cite{mor87}
These materials are cubic and the three Fourier 
components $\vec{m}_{{\vec q}_i}$ point along mutually 
perpendicular axes leading to the condition of 
a uniform magnitude of the moment.\cite{ama95}

For a triple-$\vq$ structure in UPt$_3$,
in order to explain the vanishing intensity at 
the (1/2,0,0) Bragg point we are required to have 
$\vec{m}_{\vec{q}_1}$ parallel to $\vec{q}_1$ and by symmetry the
other two moments must also be parallel to their propagation vectors.
Thus, the magnetic moment of both U ions in the n$^{\mbox{\small th}}$ 
unit cell is given by
\begin{equation}
\vec{m}_{n}=|m|\sum_{i=1}^3\,\hat{\vq}_i\,e^{i(\phi_i-\vec{\vq}_i\cdot\vR_n)}
\,.
\end{equation}

\begin{figure}
\begin{minipage}{\linewidth}
\centerline{\epsfxsize 5cm\epsffile{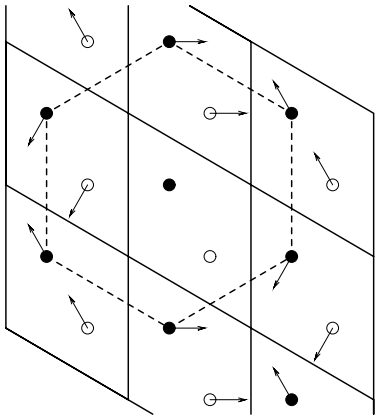}}
\vspace{0.1in}
\caption[]
{Spatial distribution of the magnetic moments for a
triple-{\bf q} magnetic structure with equal values of the
three phase factors, $\phi_i$. Note that the two U ions in the
center of the cell have zero net moment. Black filled circles represent
U atoms in the $z=c/4$ plane, and empty circles represent U
atoms in the $z=3 (c/4)$ plane.}
\label{fig:celltrq} 
\end{minipage}
\end{figure}

It can be easily shown that it is not possible to satisfy the 
condition of equal magnitude of the moment at every U site. 
Most choices for the phases $\phi_1,\phi_2,\phi_3$ 
produce a non-uniform
distribution of the magnitude of the U magnetic moment. \cite{singleq}
For example, Fig. \ref{fig:celltrq} displays a possible spatial 
distribution of the moments. The three Fourier components of the triple
structure have been chosen with equal phase $\phi_1=\phi_2=\phi_3$. 
The magnetic unit cell is then constructed from four unit cells 
containing eight U ions, reducing the hexagonal symmetry to monoclinic.
Note that the two U ions in the central cell have zero net
moment, while the other six U ions have equal values for the 
magnitude of the moment. 
 %~~~~~~~~~~~~~~~~~~~~~~~~~~~~~~~~~~~~~~~~~~~~~~~~~~~~~~~~~~~~~~~~~~~~~~~~~~~~~~
%~~~~~~~~~~~~~~~~~~~~~~~~~~~~~~~~~~~~~~~~~~~~~~~~~~~~~~~~~~~~~~~~~~~~~~~~~~~~~~~

Even though a triple-{\bf q} magnetic structure in UPt$_3$
is compatible with the neutron-scattering 
experiments
the resulting non-uniform magnetization is unusual, but not unique.
The triple-{\bf q} magnetic structure in UPt$_3$ is similar
to the magnetically frustrated structure of the uranium intermetallic
$UNi_4B$, which also has a hexagonal crystal lattice.\cite{men94}
This material orders antiferromagnetically around $T_N=30K$, with
approximately $1/3$ of the U spins remaining paramagnetic well below $T_N$.
It has been suggested that the competition between the Kondo effect, 
the antiferromagnetic exchange interaction and the frustration of 
the crystallographic lattice is responsible for the unusual $UNi_4B$
magnetic structure.\cite{lac96} Such an interplay between competing 
interactions could also take place in UPt$_3$. However, to our 
knowledge, there is no other indication of such a frustrated 
magnetic structure in $UPt_3$.

Note that a triple-$\vq$ structure does not preclude the coupling 
between the AFM  and superconducting order parameters, which is considered
a good candidate for the proposed SBF in the
2D order parameter models for the superconducting 
phases.\cite{hes89,mac89,joy92,sau94,mac98,hef96} 
The SBF coupling is non-vanishing
for triple-$\vq$ structures, except for the special case in which all 
three phases are identical. The coupling between the superconducting,
$\vec{\eta}=(\eta_1,\eta_2)$, and the magnetic order parameters is 
$F_{\mbox{\tiny AFM-SC}}\propto
  A (|\eta_1|^2 - |\eta_2|^2) + 
   B  (\eta_1\eta^{*}_2 + \eta^{*}_1\eta_2)$,
with 
$A=\sum_{n=1,4}(m^2_x(n)-m^2_y(n))= 
   4-2\cos^2(\phi_2-\phi_1)-2\cos^2(\phi_3-\phi_1)$,
$B=2 \sum_{n=1,4}(m_x(n)m_y(n))= 
   2\sqrt{3}(\cos^2(\phi_2-\phi_1)-\cos^2(\phi_3-\phi_1))$,
where the summation refers to the four unit cells contained in
the magnetic unit cell shown in Fig. \ref{fig:celltrq}.

The hexagonal triple-{\bf q} shown in Fig. \ref{fig:celltrq}
resembles the antiferroquadrupolar order reported for 
$UPd_3$.\cite{wal94,mcE95} 
Furthermore, $Pt$ and $Pd$ are isoelectronic, their nearest neighbor
U-U distances are almost identical, and both systems have a hexagonal 
closed packed structures. However, the magnetic and electronic properties 
of $UPt_3$ and $UPd_3$ are very different. In fact $UPd_3$ is a 
localized material\cite{bae80} with well-defined crystal-field 
levels.\cite{buy85} Several measurements on $UPd_3$ show {\sl two} phase 
transitions at $7 K$ and $5K$.\cite{and78,ott80} The transition at
$7K$ is believed to correspond to a 
quadrupolar ordering of the U ions, which is
accompanied by a modulated lattice distortion. The $5K$ transition is
magnetic, with an ordered moment that is very small, as in 
UPt$_3$, $\mu\simeq 0.01 \mu_B$/U-ion. But, the moments in $UPd_3$ are
pointing out of the basal plane.\cite{ste92}

We conclude with a brief discussion of possible neutron scattering 
experiments which might clarify the magnetic order in UPt$_3$.
A zero-field systematic measurement of the intensity of a number of
magnetic peaks in the same single crystal will determine whether the
magnetic moments are indeed parallel to the propagation vector or 
not. Using previous experimental arrangements\cite{lus96a} it would be
very interesting to apply fields well above $3\,T$ and measure the 
intensity at three independent momentum transfers.
Although  polarized inelastic neutron-scattering 
experiments have been performed in $UPt_3$,\cite{gol87} the
magnetic Bragg peaks have not been studied with polarized
neutrons. Polarized elastic neutron-scattering would provide 
confirmation of the magnetic nature of the transition. 
This powerful method has been used successfully  
on $UPd_3$ to identify the magnetic nature of the second phase
transition at $T_2=5 K$.\cite{ste92} 

In summary, based on available neutron diffraction data,
the magnetic field dependence of the neutron scattering intensity
is consistent with antiferromagnetic order 
in $UPt_3$ based on the most conventional assumption of a 
single-{\bf q} structure with three equivalent domains. 
However, a triplet-{\bf q} structure is also consistent with these 
experiments. If realized the triple-$\vq$ structure would imply a 
non-uniform, frustrated magnetic structure in the crystal.

We thank Piers Coleman, Bill Halperin and Robert Joynt
for valuable discussions on this subject. 
The hospitality of the Aspen Center for Physics during the 1999
Summer workshop on unconventional order in metals is gratefully acknowledged.
This research was 
supported by NSF grants DMR 9705473, DMR 9972087 and DMR 91-20000 
through the Science and Technology Center for Superconductivity.

\end{multicols}

\begin{thebibliography}{10}
\bibitem{superexchange}
At each Uranium site three mirror planes, each of them including one
of the nearest-neighbor U ions, intersect. Thus, $\vec{D}_{ij}$ for
nearest-neighbor sites must be perpendicular to the mirror plane
including $i$ and $j$. As consequence, $\vec{D}_{ij}$ would lie in
the basal plane, which contradicts the constraint $\vec{D}_{ij}||\vc$ 
imposed by the hypothesis of magnetic moments confined to the basal plane.
However, when $i$ and $j$ are next-near-neighbor U sites the only restriction
is that $\vec{D}_{ij}$ must be invariant under $\pi/3$ rotations in the
basal plane.
\end{thebibliography}
\end{document}